\documentclass[aps,preprint,showpacs,superscriptaddress]{revtex4}
\usepackage{amssymb}

\usepackage{graphicx}

\begin{document}
\title{ General theory of decoy-state quantum cryptography with source errors}
\author{Xiang-Bin Wang}
\affiliation{Department of Physics, Tsinghua University, Beijing
100084, China} \affiliation{Imai-Project, ERATO-SORST, JST, Daini
Hongo White Building, 201, 5-28-3, Hongo, Bunkyo, Tokyo 113-0033,
Japan }
\author{Cheng-Zhi Peng}
\affiliation{Department of
Physics, Tsinghua University, Beijing 100084,
China}\affiliation{Hefei National Laboratory for Physical Sciences
at Microscale and Department of Modern Physics, University of
Science and Technology of China, Hefei, Anhui 230026, China}
\author{Jun Zhang}
\affiliation{Hefei National Laboratory for Physical Sciences at Microscale and
Department of Modern Physics, University of Science and Technology
of China, Hefei, Anhui 230026, China}
\author{Lin Yang }
\affiliation{Department of Physics, Tsinghua University, Beijing
100084, China} \affiliation{Key Laboratory of Cryptologic Technology
and Information Security, Ministry of Education, Shandong
University, Jinan, China}
\author{Jian-Wei Pan}
\affiliation{Department of Physics, Tsinghua University, Beijing 100084,
China}
\affiliation{Hefei National Laboratory for Physical Sciences
at Microscale and Department of Modern Physics, University of
Science and Technology of China, Hefei, Anhui 230026,
China}
\affiliation{Physikalisches Institut, Universit\"at
Heidelberg, Philosophenweg 12, 69120 Heidelberg, Germany}

\begin{abstract}
The existing theory of decoy-state quantum cryptography assumes the
exact control of each states from Alice's source. Such exact control
is impossible in practice.  We
 develop the theory of decoy-state method so that it is
unconditionally secure even there are state errors of sources, if
the range of a few parameters in the states are known. This theory
simplifies the practical implementation of the decoy-state quantum
key distribution because the unconditional security can be achieved
with a slightly shortened final key, even though the small errors of
pulses are not corrected.
\end{abstract}


\maketitle


\section{ Introduction}  Most of the existing set-ups of quantum
key distribution (QKD)\cite{BB84,bruss,GRTZ02,DLH06,GLLP04} use
imperfect single-photon source. Such an implementation in principle
suffers from the photon-number-splitting attack \cite{PNS,PNS1}. The
decoy-state method \cite{rep, H03, wang05, LMC05,HQph} and some
other methods \cite{scran,kko,zei} can be used for unconditionally
secure QKD even Alice only uses an imperfect source\cite{PNS,PNS1}.

 The separate theoretical
result of ILM-GLLP \cite{GLLP04} shows that a secure final key can
be distilled even though an imperfect source is used in the
protocol, if one knows  the lower bound of the fraction of
 those raw bits generated by single-photon pulses
from Alice. The decoy-state method is to verify such a bound
faithfully and efficiently. The ILM-GLLP theory does not need the
{\em exact} value of the fraction of raw bits due to single-photon
pulses from Alice. It only needs the lower bound of fraction of
un-tagged bits. The goal of decoy-state method is to verify such a
lower bound through the observed experiment data.

Recently, a number of experiments on the decoy-state QKD have been
done \cite{Lo06,peng}.  However, the existing decoy-state theory
assumes the perfect control of the source states in the photon
number space. This is an impossible task for any real set-up in
practice. A new problem arose in practice is how to carry out the
decoy-state method securely and efficiently given the inexact
control of the source.
     Even though one can control
the pulse intensity pretty well in practice, we still need
quantitative criteria on the effects of possible small errors in the
source states. By currently existing technology, the source error
can be polynomially small rather than exponentially small. One may
argue sine the source error is very small,  the error-free
decoy-state theory must work. But we never know how small is
sufficiently small so that the error can be securely regarded as 0.
If we judge $20\%$ of intensity error is too large, we have no
reason to say $1\%$ error is small enough for unconditional security
with error-free decoy-state theory. There are also problems in
evaluating the different set-ups.  Suppose there are two set-ups, A
and B, they can make QKD for the same distance. Set-up A can have a
key rate of 100 bits per second with possible intensity error of
$3\%$ while set-up B can have a key rate of 80 bits per second with
possible intensity error of $1\%$, we don't know which one is
better. To answer all these questions, we need a more general theory
that directly applies to the case with state errors.
If  we have a stable two-value attenuator, we can use the method in
Ref.\cite{wangapl}. If the parameter errors of states of each pulses
are random and independent,
    we can apply the existing decoy-state method with the averaged-state\cite{wang07,HQph}.

Here we study the decoy-state method with state errors of Alice's
source without any of the above presumed conditions.
Our result
here is not limited to the intensity error only, it applies to the
more general case of state errors from the source. For example, in
the protocol using coherent states, intensity error is only a
special type of error which changes one parameter in the coherent
state only, but the state is still a coherent state, i.e., a
Possonian distribution with another parameter. Generally speaking,
there could be certain types of sour errors with which the state is
{\em not} a coherent state, i.e., not in the Possonian distribution.
Our study shows that the decoy-state method is still secure even
with state errors, but the key rate will be decreased. Our result
only needs the range of a few parameters in the source states,
regardless of whatever error pattern.  In our method, we have
assumed the worst case that Eavesdropper (Eve.) knows exactly the
error of each pulse. Our result immediately applies to all existing
experimental results.

This paper is arranged as the following. After the introduction
above, we review the main idea and basic assumptions of the existing
error-free decoy-state theory. As a result,  the most general
condition and formula for the error-free decoy-state method are
given. We then  study the main problem of this paper: the
decoy-state method with state errors from the source. After pointing
out the consequence of the source errors, we present a general
formula where only the bound values of a few parameters in the
states are involved. The result is secure even in the case Eve.
knows exactly the error of each pulse. We then point out that our
theory can be applied in the practical set-ups with state errors,
such as the decoy-state Plug-and-Play protocol raised by Gisin
group\cite{gisind}. The paper is ended with a concluding remark.
\section{  Review of the existing error-free decoy-state theory}
We call the existing
theory\cite{rep,H03,wang05,LMC05,pdc1,pdc2,haya}  {\em error-free}
theory because it assumes no error for the source states in the
photon-number space.
 Although most of the literatures\cite{rep,H03,wang05,LMC05} study the error-free decoy-state method only
 with specific distributions of source states, the assumption of a
specific distribution is actually not necessary. Basically, we only
need a few conditions for the source states rather than the complete
distribution function. Here we summarize the existing theory in the
extended format.

 In the  3-intensity protocol,  Alice has three
sources, source $Y_0$ which can produce vacuum only, source $Y$
which can only produce state
\begin{equation}\label{decoyd}
\rho=\sum_{k=0}^J a_k|k\rangle\langle k|
\end{equation}
only, and source $Y'$ which can only produce state
\begin{equation}\label{signald}
\rho'=\sum_{k=0}^J a_k'|k\rangle\langle k|
\end{equation}
only, where $|k\rangle$ is the $k-$photon Fock state and $a_k\ge
0,\;a_k'\ge 0$ for all $k$, $\sum a_k=\sum a_k'=1$. Here $J$ can be
either finite or infinite. Given a coherent-state source or a
heralded single-photon source from the parametric down conversion,
$J=\infty$. When a coherent-state source\cite{rep,H03,wang05,LMC05}
or a heralded single-photon source\cite{pdc1,pdc2} is used, the
parameters $a_k,a_k'$ are determined by the intensity (averaged
photon number) of a pulse.

For simplicity, we shall also call source $Y_0,\;Y,\;Y'$ the vacuum
source, the decoy source, and the signal source, respectively.
Pulses from  the decoy source or the signal source are called {\em
the decoy pulses} or {\em the signal pulses}, respectively. In the
protocol, Alice may use each source of $\{Y_0,Y,Y'\}$ randomly with
probabilities $p_0,p,p'$ ($p_0+p+p'=1$) whenever she sends out a
pulse to Bob. Given the states in Eqs.(\ref{decoyd},\ref{signald}),
we can equivalently regard that source $Y$ or $Y'$ as a
probabilistic photon-number source which sends out a {\em k-}photon
pulse (photon number state $|k\rangle$) with probability
distribution $\{a_k\}$ or $\{a_k'\}$. The task here is to verify the
lower bound of the raw bits caused by those single-photon pulses
from Alice.  Alice and Bob can not directly observe the number of
single-photon counts because they don't know which pulses at Alice's
side are single-photon pulses. They only know which pulse belongs to
which source and the states of  each source. The decoy-state theory
shows that they only need to know the number of counts due to the
pulses from each sources in order to verify how many  raw bits are
generated from the single-photon pulses from Alice.

We first define the {\em counting rate} of a class (or a sub-class)
of pulses. A class or a sub-class can be any set or subset of pulses
from Alice.  In the 3-intensity decoy-state method, Alice has 3
sources. Here we shall regard each source as a class, and all those
{\em k-}photon ($k=0,1,2,\cdots$) pulses from the same source as a
subclass. Given any class $X$ that contains $M_X$ pulses, after
Alice sends them out to Bob, if Bob observes $n$ counts at his side,
the counting rate for pulses in this class is
\begin{equation}S_x=n/M_X.\label{def}\end{equation}
If class $X$ is divided into $J$ sub-classes and any pulse in class
$X$ belongs to and only belongs to one subclass and the fractions of
pulses in each subclasses are $b_0,b_1\cdots,b_J$, then the counting
rate of class $X$ is
\begin{equation}\label{sum} S_X=\sum_{k=0}^J b_ks_k \end{equation}
and $s_k$ is the counting rate of the $k$th subclass. This is simply
because the total counts of a class equals to the summation of
counts of each sub-classes, i.e.,
\begin{equation}\label{sumc}
n=\sum_{k=0}^J n_k
\end{equation}
and $n_k$ is the number of counts at Bob's side caused by pulses in
the $k$th sub-class from Alice. We denote the counting rates of the
decoy-source (class $Y$), the signal source (class $Y'$), and the
vacuum source (class $Y_0$) by $S,\;S',$ and $S_0$, respectively.
Since in the protocol, Alice knows which pulse is from which source,
these $S,\;S',$ and $S_0$ can be observed directly in the experiment
therefore we regard them as {\em known} parameters. Given the state
of the decoy pulses in Eq.(\ref{decoyd}), the fraction of $k-$photon
pulses is $a_k$. Then we have
\begin{equation}\label{countdecoy}
S = \sum_{k=0}^J a_ks_k=a_0S_0+a_1s_1+\lambda
\end{equation}
and $s_k$ is the counting rate of those $k-$photon pulses from
source $Y$,
and\begin{equation}\lambda=\sum_{k=2}^{J}a_ks_k .\end{equation}

Similarly, we also have
\begin{equation}\label{countsignal}
S' = \sum_{k=0}^J a_k's_k'
\end{equation}
and $s_k'$ is the
counting rate of  those $k-$photon pulses from source $Y'$, $S'$ is the counting rate
of all pulses from source $Y'$.

Asymptotically, the counting rate of the sub-class of those {\em
k-}photon pulses from the decoy source  and the sub-class of those
{\em k-}photon pulses from the signal source must be equal
\begin{equation}\label{0011cc}s_k=s_k'
\end{equation}
{\em if} the pulses from these two subclasses are {\em randomly
mixed}.
 According to this, we also have,
 \begin{equation}\label{0011cc0}
s_0=s_0'=S_0,
\end{equation}
 Here $S_0$ is the counting rate
of class $Y_0$. In the protocol Alice randomly uses three sources
therefore those {\em k-}photon pulses from the decoy-source and
those {\em k-}photon pulses from the signal source are randomly
mixed. We can rewrite Eq.(\ref{countsignal}) in the format
\begin{equation}\label{cs1}
S'=\sum_{k=0}^{J}a_k's_k=
a_0's_0+a_1's_1+\frac{a_2'}{a_2}\lambda +\delta
\end{equation}
and
\begin{equation}
\delta =\sum_{k=2}^J a_k's_k-\frac{a_2'}{a_2}\lambda.
\end{equation}
Obviously,
\begin{equation}
\delta\ge 0
\end{equation}
if
\begin{equation}\label{ac}
\frac{a_k'}{a_k}\ge \frac{a_2'}{a_2}\ge\frac{a_1'}{a_1}.
\end{equation}
Given Eq.(\ref{countdecoy},\ref{cs1}), one can find the following fact
\begin{equation}
s_1=s_1'= \frac{a_2'( S-a_0s_0)-a_2 (
S'-a_0's_0)}{a_2'a_1-a_1'a_2}+\frac{a_2\delta}{a_2'a_1-a_1'a_2}.
\end{equation}
Since $\delta\ge 0$, and the second inequality in Eq.(\ref{ac}) is
\begin{equation}\label{ac1}
a_2'a_1-a_1'a_2\ge 0,
\end{equation}
 therefore $\frac{a_2\delta}{a_2'a_1-a_1'a_2}\ge 0$. The minimum value of the single-photon counting rate is
now verified to be
\begin{equation}\label{f11}
s_1=s_1'\ge \frac{a_2'( S-a_0s_0)-a_2 (
S'-a_0's_0)}{a_2'a_1-a_1'a_2}.
\end{equation}
The fractions of the single-photon counts for the signal source and
the decoy source are therefore
\begin{equation}\Delta_1'=\frac{a_1's_1}{S'},\;\Delta_1=\frac{a_1s_1}{S}.\end{equation}
 Given these, one can calculate the final key rate of each source, e.g., the final
 key rate for the signal source is
by\cite{GLLP04,LMC05}
\begin{equation}\label{ilm}
R_s=\Delta_1'[1-H(t_1)]-H(t)
\end{equation}
where
 $t_1$, $t$ are the QBER for single-photon pulses and the QBER for
all signal pulses. This is the (extended) result of the decoy-state
method with diagonal states in photon number space, including the
coherent states, thermal states, heralded single-photon states, and
so on with the condition that the source states are exactly
controlled and Eq.(\ref{ac}) holds.

\section{Consequence of source errors: $s_k\not=s_k'$}
However, the results above are based on the assumption that all
states are produced exactly. Now we study the the consequence of the
source errors in an actual protocol.

A very tricky point here is that Eq.(\ref{0011cc}) is in general
incorrect, if there are state errors. We emphasize that this issue
was first pointed out by us in a number of places. For example, in
the shorter version of our work \cite{note} (in the text around
Eq.(11) there), and also  in  section 4.3.5 of Ref.\cite{rep}. In
all these places we have clearly stated the issue. Here we present
the main idea about this point and also the reason behind:

Eq.(\ref{0011cc}) is the most important element of the error-free
decoy-state theory. Now we explain why we don't use
Eq.(\ref{0011cc})  here if we assume Eve. knows the state errors.

 For simplicity, we consider the  following two-block
collective errors with coherent states: Alice wants to use intensity
$\mu=0.2$ for decoy pulses and intensity $\mu'=0.6$ for signal
pulses. However, in some blocks (strengthened blocks), both decoy
pulses and signal pulses are 10\% stronger than the assumed
intensity, in the other blocks (weakened blocks),  both decoy pulses
and signal pulses are 10\% weaker than the assumed values\cite{rep}.
Suppose Alice always with probabilities of $p_0,p,p'$ to choose one
of three different sources (the vacuum source, the decoy source, and
the signal source source).   The density operator of a coherent
state with intensity $x$ is
\begin{equation}
\rho_x=e^{-x}\sum_{k=0}^{\infty}\frac{x^k}{k!}|k\rangle\langle k|.
\end{equation}
Here is Eve's scheme using {\bf time-dependent channel}: she blocks
all pulses from the weakened blocks, and she produces a linear
channel of transmittance $2\eta_e$ to attenuate each pulse from the
strengthened block. Straightly, the actual counting rate of those
single-photon pulses from the decoy source is
\begin{equation}
s_1=\frac{\frac{1}{2}\times 2\eta_e\times 1.1 \mu
e^{-1.1\mu}}{(0.9\mu e^{-0.9\mu}+1.1\mu e^{-1.1\mu})/2}
\end{equation} and the
actual counting rate of those single-photon pulses from signal
source is
\begin{equation}
s_1'=\frac{\frac{1}{2}\times 2\eta_e\times 1.1 \mu'
e^{-1.1\mu'}}{(0.9\mu' e^{-0.9\mu'}+1.1\mu' e^{-1.1\mu'})/2}.
\end{equation}
We find
\begin{equation}
s_1/s_1'=\frac{e^{0.2\mu'}+1.1/0.9}{e^{0.2\mu}+1.1/0.9}\not= 1.
\end{equation}
Similarly, we can also show that $s_k\not=s_k'$ for any $k$. This
shows, given the collective error which is known to Eve, Eve can
treat the identical $k-$photon pulses {\em differently} according to
which source the are from !

The reason behind the above result is because the collective errors
may break the ``random mixture" condition of Eq.(\ref{0011cc}).
Consider again the specific example above. Consider those
single-photon pulses only. In a strengthened block, a single-photon
pulse has probability
\begin{equation}
{\mathcal P}_{s}=\frac{1.1p\mu e^{-1.1\mu}}{1.1p\mu
e^{-1.1\mu}+1.1p'\mu'e^{-1.1\mu'}}=\frac{1}{1+\frac{p'\mu'}{p\mu}e^{1.1(\mu-\mu')}}
\end{equation}
to be from the decoy source; and probability
\begin{equation}{\mathcal P}_s'= \frac{1.1p'\mu e^{-1.1\mu'}}{1.1p\mu
e^{-1.1\mu}+1.1p'\mu'e^{-1.1\mu'}}=\frac{1}{1+\frac{p\mu}{p'\mu'}e^{1.1(\mu'-\mu})}
\end{equation}
to be from the signal source. In a weakened block, a single-photon
pulse has probability
 \begin{equation}
{\mathcal P}_{w}=\frac{1}{1+\frac{p'\mu'}{p\mu}e^{0.9(\mu-\mu')}}
\end{equation}
that  it comes from the decoy source; and probability
\begin{equation}{\mathcal P}_w'=\frac{1}{1+\frac{p\mu}{p'\mu'}e^{0.9(\mu'-\mu})}
\end{equation}
that  it comes from the signal source. These values show that the
single-photon pulses in the weakened blocks are more probably from
 the decoy source than those single-photon pulses in  the strengthened blocks.
 Also, the
single-photon pulses  in the weakened blocks are less probably from
the signal source than those single-photon pulses in the
strengthened blocks. These patterns have surely broken the random
mixture presumption of Eq.(\ref{0011cc}).

One may question how Eve. can know the state errors. In the
decoy-state plug-and-play protocol\cite{gisind},  Eve. can actually
prepare the error pattern, including the above specific pattern.
(Suppose Alice only wants to monitor each pulse  but she does not
want to cost too much to correct the the error of each individual
pulse.)

Even in the one-way protocol, Eve. can also know the intensity
errors if Alice does not want to cost too much to correct each
individual pulse. Indeed, in the practical set-ups, the intensity
fluctuation appears {\em collectively} to blocks of pulses. Eve. can
know the error patterns by studying the averaged photon number block
by block.

In principle, Alice might be able to correct the error of each
individual pulses in a real set-up. But, with the theoretical result
of this work,  Alice does not have to do so. She only needs to know
the range of a few parameters of the  source.

\section{ Decoy-state method with state errors}
We still assume that each pulse sent out by Alice is randomly chosen
from one of 3 sources $Y_0,Y,Y'$ with probability $p_0,p,p'$,
respectively. For simplicity, we assume that every pulse in class
$Y_0$ is exactly in vacuum state. But each single-shot of pulses in
classes $Y$ (the decoy source), and $Y'$ (the signal source) can be
in a state slightly different from the expected one. Eq.(\ref{sumc})
still holds even though there are source errors, because it is
simply the definition of summation. Also, we shall still use
Eq.(\ref{def}) for the definition of  counting rate of a certain
class which contains many pulses.  Suppose Alice sends $M$ pulses to
Bob in the whole protocol.
\subsection{Virtual protocol}
For clarity, we
 first
consider a virtual protocol,\\ {\bf Protocol 1}: At any time $i$,
each source produces a pulse. The states of the pulses from sources
$Y_0,Y\;,Y'\;$ are $|0\rangle\langle 0|$,
\begin{equation}\label{rhoi}\rho_i=\sum_{k=0}^J a_{ki}|k\rangle\langle k|\; ;\;{\rm and}\end{equation}
\begin{equation}\label{rhoip} \rho_i'=\sum_{k=0}^J a_{ki}'|k\rangle\langle k|.\end{equation}
Here $\rho_i$, and $\rho_i'$ can be a bit different from $\rho$ and
$\rho'$ of Eq.(\ref{decoyd}, \ref{signald}), which are the assumed
states in the perfect protocol where there is on source error. At
any time $i$, only one pulse is selected and sent out for Bob, and
the probability for a pulse to be selected and sent out is
constantly $p_0,\;p$, and $p'$ if the pulse is from vacuum source
(decoy-source or signal source). The un-selected two pulses at each
time will be blocked and absorbed. After Bob has completed all
measurements to the incident pulses, Alice checks the record about
which pulse is selected at each time, i.e., which time has used
which source.

As shown below, based on this virtual protocol, we can formulate the
number of counts from each source and therefore find the lower bound
of the number of single-photon counts. The result also holds for the
real protocol where Alice decides to use which sources at the $i$th
time in the beginning with the same probability distribution
$p_0,\;p$ and $p'$. In this case, at each time $i$, only one source
emits a pulse.
\subsection{Our goal}\label{goal} Our {\em  goal} is to
find the  lower bound of the fraction of counts caused by those
single-photon pulses for both the signal source and the decoy
source. The following quantities  are directly observed in the
protocol therefore we regard them as known parameters: $N_d$, the
number of counts caused by the decoy source, and $N_s$, the number
of counts caused by the signal source, and $N_0$, the number of
counts caused by the vacuum source, $Y_0$. Obviously,
$S=\frac{N_d}{pM}$, the counting rate of the decoy source, and
$N_s=\frac{N_s}{p'M}$, the counting rate of the signal source, and
$S_0=\frac{N_0}{p_0M}$, the counting rate of the vacuum source, are
also known exactly in the protocol. Clearly, here $pM$, $p'M$ and
$p_0M$ are just the number of the decoy pulses, the number of the
signal pulses, and the number of pulses from the vacuum source,
respectively.  Therefore, for our goal, we only need to formulate
the number of counts caused by those single-photon pulses from each
sources in terms of these quantities and $p_0,p,p'$ and the bound
values of those parameters $a_{ki},a_{ki}'$ as appear in
Eq.(\ref{rhoi},\ref{rhoip}).
\subsection{Some definitions}
{\em Definition 1}. In the protocol, Alice sends Bob $M$ pulses, one
by one. In response to Alice, Bob observes his detector for $M$
times. As Bob's {\em i}th observed result, Bob's detector can either
click or not click. If the detector clicks in Bob's {\em i}th
observation, then we say that ``the {\em i}th pulse from Alice has
caused a count". We disregard how the {\em i}th pulse may change
after it is sent out. When we say that Alice's {\em i}th pulse has
caused a count we only need Bob's detector clicks in Bob's {\em i}th
observation.

Given the source state in Eqs.(\ref{rhoi},\ref{rhoip}), any $i$th
pulse sent out by Alice must be in a photon-number state. To anyone
outside Alice's lab, it looks as if that Alice only sends a photon
number state at each single-shot: sometimes it's vacuum, sometimes
it's a single-photon pulse, sometimes it is a $k-$photon pulses, and
so on. We shall make use of this fact that any individual pulse is
in one Fock state.
\\
{\em Definition 2}, set $C$ and $c_k$:  Set $C$ contains any {\em
B-}pulse that has caused a count; set $c_k$ contains any $k-$photon
{\em B-}pulse that has caused a count. Mathematically speaking, the
sufficient and necessary condition for $i\in C$ is that  the {\em
i}th pulse has caused a count.  The sufficient and necessary
condition for $i\in c_k$ is that  the {\em i}th pulse contains $k$
photons and it has caused a count. For instance, if the photon
number states of the first 10 pulses from Alice are
$|0\rangle,\;|0\rangle,\;|1\rangle,\;|2\rangle,\;|0\rangle,\;
|1\rangle,\;|3\rangle,\;|2\rangle,\;|1\rangle,\;|0\rangle,\;$  and
the pulses of $i=2,\; 3,\; 5,\; 6,\; 9,\; 10$ each has caused a
count at Bob's side, then we have
\begin{equation}
C=\{i|i=2,\;3,\;5,\;6,\;9,\;10,\cdots\};\;c_0=\{i|i=2,5,10,\cdots\};
\; c_1=\{i|i=3,6,9,\cdots\}.
\end{equation}
Clearly, $C=c_0\cup c_1\cup c_2\cdots$, every pulse in set $C$ has
caused a count.
\\{\em Definition 3}. For any $k\ge 0$, for the parameters  in
Eqs.(\ref{rhoi}, \ref{rhoip}), we denote $a_k^L$ and $a_k^U$  the
minimum value and maximum value of $\{a_{ki}|\;i=1,2,\cdots M\}$;
$a_k'^L$ and $a_k'^U$ the minimum value and maximum value of
$\{a_{ki}'|\;i=1,2,\cdots,M\}$. We assume these bound values are
known in the protocol.
\subsection{Number of vacuum counts}
Here we shall give the explicit formulas to bound $n_{0d}$, the
number of counts caused by those vacuum pulses from the decoy
source, and $n_{0s}'$, the number of counts caused by those vacuum
pulses from the signal source. We want to formulate them by  the
known quantities, such as $N_0$, the number of counts by the vacuum
source (source $Y_0$), or $S_0$, the counting rate of the vacuum
source.

Given the definition of the set $c_k$, the number of counts caused
by all vacuum pulses is just the number of pulses in set $c_0$. We
want to know how many of the vacuum counts  are caused by each
source. This is equivalent to ask how many of pules in set $c_0$
come from each source.
 A vacuum pulse can come from any
of  the 3 sources, the vacuum source ($Y_0$), the decoy source ($Y$)
and the signal source ($Y'$).  According to Eqs.(\ref{rhoi},
\ref{rhoip}), if the {\em i}th pulse is vacuum,  the probability
that it comes from  the vacuum source ($Y_0$) is
$$\mathcal P_{vi|0}=\frac{p_0}{p_0+pa_{0i}+p'a_{0i}'}.$$
Asymptotically, in set $c_0$, the population of pulses from source
$Y_0$ is
$$ \sum_{i\in c_0}\mathcal P_{vi|0}.$$
 This is also the number of counts caused by source $Y_0$, since
 every pulse in $c_0$ has caused a count.
 Therefore the observed
number of counts caused by source $Y_0$ must satisfy
\begin{equation}\label{vcct}
N_0 = \sum_{i\in c_0} p_0d_{0i}
\end{equation}
Here
\begin{equation}
 d_{0i}=\frac{1}{p_0+pa_{0i}+p'a_{0i}'}.
\end{equation}
Similarly, if the {\em i}th  pulse contains 0 photon, it has a
probability $ \mathcal
P_{di|0}=\frac{pa_{0i}}{p_0+pa_{0i}+p'a_{0i}'}$ to be from the decoy
source, and a probability of $ \mathcal
P_{si|0}=\frac{p'a_{0i}'}{p_0+pa_{0i}+p'a_{0i}'}$ to be from the
signal source. Therefore we have
\begin{equation}\label{s0d2}
n_{0d} =  \sum_{i\in c_0} \mathcal P_{di|0}=\sum_{i\in
c_0}pa_{0i}d_{0i},
\end{equation}
for the number of counts caused by those vacuum pulses from the
decoy source, and
\begin{equation}\label{s0s2}
n_{0s}' =\sum_{i\in c_0} \mathcal P_{si|0}= \sum_{i\in
c_0}p'a_{0i}'d_{0i} ,
\end{equation}
for the number of counts caused by those vacuum pulses from the
signal source.
 Therefore, with our {\em Definition 3}, $n_{0d},\;n_{0s}'$ are bounded by
\begin{equation}
n_{0d}^U= pa_0^U \sum_{i\in c_0} d_{0i} \ge  n_{0d} \ge  pa_0^L
\sum_{i\in c_0} d_{0i} = n_{0d}^L
\end{equation}
and
\begin{equation}
n_{0s}'^U=p'a_0'^U\sum_{i\in c_0} d_{0i}\ge n_{0d}\ge
p'a_0'^L\sum_{i\in c_0} d_{0i}=n_{0s}'^L.
\end{equation}
Using the fact  $\sum_{i\in c_0} d_{0i}=\frac{N_0}{p_0}$ from
Eq.(\ref{vcct}), we replace the above two equations by
\begin{eqnarray}\label{vcbound}
n_{0d}^U=\frac{pa_0^UN_0}{p_0}\ge n_{0d}\ge \frac{pa_0^LN_0}{p_0}=n_{0d}^L\nonumber\\
n_{0s}'^U=\frac{p'a_0'^UN_0}{p_0}\ge n_{0s}'\ge
\frac{p'a_0'^LN_0}{p_0}=n_{0s}'^L
\end{eqnarray}
 By definition, the counting rate of source $Y_0$ is $S_0=\frac{N_0}{p_0M}$, therefore the above equations
  are equivalent to
 \begin{eqnarray}\label{vcbeq}
n_{0d}^U=a_0^UpS_0M\ge n_{0d}\ge a_0^LpS_0M=n_{0d}^L\nonumber\\
n_{0s}'^U=a_0'^Up'S_0M\ge n_{0s}'\ge a_0'^Lp'S_0M=n_{0s}'^L.
\end{eqnarray}
\subsection{Calculations and main formulas}
According to
our definition  of set $C$ earlier, every pulse in set $C$ has
caused a count. Therefore, the population of the decoy pulses
(signal pulses) in set $C$ is just $N_d$ (or $N_s$), the number of
counts of  the decoy source (signal source). Asymptotically,
\begin{equation}\label{popd}
N_d=\sum_{k=0}^J\sum_{i\in c_k}\mathcal P_{di|k}=n_{0d}+\sum_{i\in
c_1}\mathcal P_{di|1}+\sum_{k=2}^J\sum_{i\in c_k}\mathcal P_{di|k}
\end{equation}
\begin{equation}\label{pops}
N_s=\sum_{k=0}^J\sum_{i\in c_k}\mathcal P_{si|k}=n_{0s}'+\sum_{i\in
c_1}\mathcal P_{si|1}+\sum_{k=2}^J\sum_{i\in c_k}\mathcal P_{si|k}
\end{equation}
and $\mathcal P_{di|k}$ (or $\mathcal P_{si|k}$ )is the probability
that the {\em i}th pulse comes from the decoy source (or signal
source); if the {\em i}th pulse contains {\em k} photons. Here we
have used Eqs.(\ref{s0d2}, \ref{s0s2}).

 Consider those {\em k-}photon pulses ($k\ge 1$).
 A
{\em k-}photon pulse can come from either the decoy source  or the
signal source.
\\ {\em Fact :} Define
\begin{equation}\label{dki}
d_{ki}=\frac{1}{pa_{ki}+p'a_{ki}'},\;{\rm for}\; k\ge1,
\end{equation}
if  the $i$th pulse contains $k$ photons, it has a probability
$pa_{ki}d_{ki}$ to be from the decoy source (source $Y$), and a
probability $p'a_{ki}'d_{ki}$
 to be from the signal source (source $Y'$),  if $k\ge 1$.
This is to say, $\mathcal P_{di|k}$ in  Eq.(\ref{popd}) and
$\mathcal P_{si|k}$ in  Eq.(\ref{pops})are given by
\begin{equation}\label{popdk}
 \mathcal P_{di|k}=pa_{ki}d_{ki}\end{equation}
 and
 \begin{equation}\label{popsk}
 \mathcal P_{si|k}=p'a_{ki}'d_{ki}\end{equation}
Therefore,  Eqs.(\ref{popd}, \ref{pops}) can be re-written in the
following equivalent form
\begin{equation}\label{popde}
N_d= n_{0d}+p \sum_{i\in c_1}a_{1i}d_{1i}+p\sum_{k=2}^J\sum_{i\in
c_k}a_{ki}d_{ki},\end{equation}
\begin{equation}\label{popse}
N_s= n_{0s}'+p'\sum_{i\in c_1}a_{1i}'d_{1i}+p'\sum_{k=2}^J\sum_{i\in
c_k}a_{ki}'d_{1i}.\end{equation}

 Our goal as stated in the subsection \ref{goal}  is simply to know the minimum value of
 \begin{equation}\label{D1}
D_1=\sum_{i\in c_1} d_{1i}.
\end{equation}
 For, with this and {\em Definition 3}, the minimum value of the number of counts caused by
single-photon pulses from the signal-source (or the decoy-source) is
simply
\begin{equation}\label{n1ds}n_{1s}'^L= p'a_{1}'^LD_1\le n_{1s}',\; {\rm (or\;} n_{1d}^L
=pa_1^LD_1\le n_{1d}{\rm )}.
\end{equation}  In what follows we shall find the formula of
$D_1$ in terms of $N_d,N_s,n_{0d},n_{0s}'$ based on
Eqs.(\ref{popde}, \ref{popse}). [$n_{0d},n_{0s}'$ have been given in
Eq(\ref{vcbound})]. Eqs.(\ref{popde}, \ref{popse}) can be written in
\begin{eqnarray}\label{d0d}
N_d=
n_{0d}+pa_{1}^UD_1 +p\Lambda -\xi_1\\
N_s=n_{0s}'+ p'a_{1}'^LD_1 +p'\Lambda' +\xi_2 \label{s0s}\end{eqnarray}
where
\begin{equation}
\Lambda=\sum_{k=2}^J a_{k}^U\sum_{i\in c_k} d_{ki};\;
\Lambda'=\sum_{k=2}^J a_{k}'^L\sum_{i\in c_k}d_{ki},
\end{equation}
and
  \begin{eqnarray}\xi_1=p\left[a_{1}^UD_1 +\Lambda-
\left(\sum_{i\in c_1} a_{1i}d_{1i} +\sum_{k=2}^J\sum_{i\in c_k} a_{ki}d_{ki}\right)\right]\ge 0\nonumber\\
\xi_2=p'\left[\sum_{i\in c_1} a_{1i}'d_{1i} +\sum_{k=2}^J \sum_{i\in
c_k}a_{ki}'d_{ki}- \left( a_{1}'^L D_1 + \Lambda'\right)\right]\ge
0\nonumber
\end{eqnarray}
According to the definition of $\Lambda$ and $\Lambda'$, we also have
\begin{equation}\label{lambda}
\Lambda'=\frac{a_2'^L}{a_2^U}\Lambda + \xi_3
\end{equation}
and \begin{equation}\xi_3= \Lambda'-\frac{a_2'^L}{a_2^U}\Lambda \end{equation}
Further, we assume the important condition
\begin{equation}\label{ace}
\frac{a_k'^L}{a_k^U}\ge  \frac{a_2'^L}{a_2^U}\ge  \frac{a_1'^L}{a_1^U},\;{\rm for\; all}\;\; k\ge 2.
\end{equation}
The first inequality above leads to
\begin{equation}
\xi_3\ge 0
\end{equation}
as one may easily prove. With Eq.(\ref{lambda}),  Eq.(\ref{s0s}) is
equivalent to
\begin{eqnarray}
N_s=n_{0s}'+ p'a_{1}'^LD_1 +p'\frac{a_2'^L}{a_2^U}\Lambda
+\xi_2+\xi_3 \label{s0s1}\end{eqnarray} Given the Eqs.(\ref{d0d},
\ref{s0s1}),
 we can formulate $D_1$:
\begin{equation}
D_1=\frac{a_2'^LN_d/p-a_2^UN_s/p'-a_2'^Ln_{0d}/p+a_2^Un_{0s}'/p'+a_2'^L\xi_1/p+a_2^U(\xi_2+\xi_3)/p'}{a_1^Ua_2'^L-a_1'^La_2^U}.
\end{equation}
Since $\xi_1,\xi_2,$ and $\xi_3$ are all non-negative, and
$a_1^Ua_2'^L-a_1'^La_2^U\ge 0$ by the second inequality of
Eq.(\ref{ace}), we now have
\begin{equation}\label{med}
D_1\ge
\frac{a_2'^LN_d/p-a_2^UN_s/p'-a_2'^Ln_{0d}/p+a_2^Un_{0s}'/p'}{a_1^Ua_2'^L-a_1'^La_2^U}.
\end{equation}
The bound values of  $n_{0d},\;n_{0d}'$ have been given by
Eq.(\ref{vcbound}, \ref{vcbeq}) in the earlier subsubsection.
 Therefore, we can now bound the fraction of single counts among all counts caused by the signal source
\begin{equation}\label{main0}
\Delta_1'\ge \frac{p'a_1'^LD_1}{N_s}\ge
\frac{a_1'^L(a_2'^LN_dp'/p-a_2^UN_s-p'a_2'^La_0^UN_0/p_0+a_2^Ua_0'^LN_0p'/p_0)}{N_s(a_1^Ua_2'^L-a_1'^La_2^U)}
.
\end{equation}
According to Eq.(\ref{n1ds}),  $p'a_1'^LD_1$ is the lower bound of
the number of counts caused by single-photon pulses from the signal
source. Here we have replaced $n_{0d}$ in Eq.(\ref{med}) by its
upper bound and $n_{0s}'$ by its lower bound as given in
Eq.(\ref{vcbound}).
 Using Eq.(\ref{vcbeq}), we can write the right-hand-side of
the inequality in terms of counting rates:
\begin{equation}\label{main1}
\Delta_1'\ge \frac{a_1'^L\left(a_2'^LS-a_2^US'-a_2'^La_0^US_0+a_2^Ua_0'^LS_0\right)}
{S'\left(a_1^Ua_2'^L-a_1'^La_2^U\right)}
\end{equation}
where $S'=\frac{N_s}{p'M}$ is the counting rate of the signal
source, $S=\frac{N_d}{pM}$ is the counting rate of the decoy source,
and $M$ is the total number of pulses as defined earlier. Similarly,
we also have
\begin{equation}\label{main2}
\Delta_1\ge \frac{a_1^L\left(a_2'^LS-a_2^US'-a_2'^La_0^US_0+a_2^Ua_0'^LS_0\right)}
{S\left(a_1^Ua_2'^L-a_1'^La_2^U\right)}
\end{equation}
for the minimum value of fraction of single-photon counts for the
decoy source.

Eqs.(\ref{ace}, \ref{main0}, \ref{main1}) and (\ref{main2}) are our
main results of this work. The results are based on  the virtual
protocol where Alice checks which source is used at each time {\em
after} Bob's detection.  Obviously Alice can choose to check the
information before sending out the pulses, i.e., Alice can decide
which source to be used at each time in the very beginning. This is
then just the real protocol of the decoy-state method.

For coherent states, if the intensity is bounded by $[\mu^L,\mu^U]$
for the decoy pulses and $[\mu'^L,\mu'^U]$ for the signal pulses
then
\begin{equation}
a_k^{X}=(\mu^X)^ke^{-\mu^X}/k!,~a_k'^{X}=(\mu'^X)^ke^{-\mu'^X}/k!
\end{equation}
with $X=L,\;U$ and $k=1,2$ and
\begin{equation}
a_0^{L}=e^{-\mu^U},\;a_0^{U}=e^{-\mu^L}\;a_0'^{L}=e^{-\mu'^U},\;a_0'^{U}=e^{-\mu'^L}
\end{equation}  Therefore,  one can calculate the final key rate  by
Eq.(\ref{ilm}) now, if the bound values of intensity errors  are
known. The asymptotic result using the experimental data of QKD over
50 kilometers calculated by our formula is listed in table I.


\begin{table}
\caption{\label{tab:table1}Secure key rate ($R$) vs different values
of intensity error upper bound ($\delta_M$) using the experimental
data in the case of 50~km~\cite{peng}. The experiment lasts for
1481.2 seconds with the repetition rate 4~MHz. We have observed
$S'=3.817\times 10^{-4}, S=1.548\times 10^{-4}, S_{0}=2.609\times
10^{-5}$ and the  quantum bit error rates (QBER) for signal states
and decoy states are $4.247\%, 8.379\%$ respectively. The relative
fractions of the signal pulses, decoy pulses and the pulses from the
vacuum source are 0.50269: 0.40726: 0.09006. The final key rate is
calculated by Eq.(\ref{ilm}) and we use the formula
$\frac{4.247\%}{\Delta_1'}$ for $t_1$ in Eq.(\ref{ilm}).}
\begin{ruledtabular}
\begin{tabular}{lllllll}
 $\delta_M$ & $5\%$ & $4\%$ & $3\%$ & $2\%$ & $1\%$ & 0\\
\hline
$R$ (Hz) & 70.8 & 84.3 & 97.6 & 110.7 & 123.6 & 136.3\\
\end{tabular}
\end{ruledtabular}
\end{table}
\section{ Application in the Plug-and-Play protocol}
 As shown by Gisin et al\cite{gisind}, combining with the decoy-state method,
 the plug-and-play protocol can be unconditionally secure.
  There, Alice receives strong pulses from Bob
and she needs to guarantee the exact intensity of the pulse sending
to Bob. It is not difficult to check the intensity, but difficult to
{\em precisely  correct} the intensity of each individual pulses.
Our theory here can help to save the difficult single-shot
feed-forward intensity control: Alice monitors each pulses, to
reduce the cost of the set-up, she may only do crude corrections to
the pulses, or she may simply discard those pulses whose intensity
errors are too large (e.g., beyond 2\%), and then use our theory
with the known bound of state errors. In the Plug-and-Play protocol,
Eve. actually knows the error of each individual pulse hence the
error-free decoy-state theory based on Eq.(\ref{0011cc}) fails but
our theory here works.
\section{ Concluding remark and discussions} In summary, we have for the first time shown  the unconditional security
of decoy-state method given what-ever error pattern of the source,
provided that the parameters in the diagonal state of the source
satisfy Eq.(\ref{ace}) and the bound values of each parameters in
the state is known.
 Our result also
answers clearly the often asked question ``What happens if the state
of Laser beam is not exactly in the assumed distribution ?". Our
result can be directly applied to the Plug-and-Play decoy-state
protocol and simplify all the existing protocols in practical use.
Our result here can be extended to the non-asymptotic case by taking
statistical fluctuations into consideration in Eqs.( \ref{popd},
\ref{pops}). This will be reported elsewhere.
 \\{\bf Acknowledgement:} This work was
supported in part by the National Basic Research Program of China
grant No. 2007CB907900, 2007CB807901, 2007CB807902 and 2007CB807903,
NSFC grant No. 60725416 and China Hi-Tech program grant No.
2006AA01Z420.


\begin{thebibliography}{99}
\bibitem{BB84}
C.H.~Bennett and
G.~Brassard, in {\em Proc.\ of IEEE Int.\ Conf.\ on Computers,
Systems, and Signal Processing (IEEE, New York, 1984)},
pp.~175-179.
\bibitem{bruss} D. Bruss, Phys. Rev. Lett, 81,
3018(1998).
\bibitem{GRTZ02}
N.~Gisin, G.~Ribordy, W.~Tittel, and H.~Zbinden, Rev. Mod. Phys.
{\bf 74}, 145 (2002).

\bibitem{DLH06}
M.~Dusek, N.~L\"utkenhaus, M.~Hendrych, in
{\em Progress in Optics VVVX}, edited by E.~Wolf (Elsevier, 2006).
\bibitem{GLLP04}
H.~Inamori, N.~L\"utkenhaus, D.~Mayers, quant-ph/0107017;
D.~Gottesman, H.K.~Lo, N.~L\"{u}tkenhaus, and J.~Preskill, Quantum
Inf. Comput. {\bf 4}, 325 (2004).

\bibitem{PNS}
B.~Huttner, N.~Imoto, N.~Gisin, and T.~Mor, Phys. Rev. A {\bf 51},
1863 (1995); H.P.~Yuen, Quantum Semiclassic. Opt. {\bf 8}, 939 (1996).

\bibitem{PNS1}
G.~Brassard, N.~L\"utkenhaus, T.~Mor, and
B.C.~Sanders, Phys. Rev. Lett. {\bf 85}, 1330 (2000);
N.~L\"utkenhaus, Phys. Rev. A {\bf 61}, 052304 (2000);
N.~L\"utkenhaus and M.~Jahma, New J. Phys. {\bf 4}, 44 (2002).
\bibitem{rep}X.-B. Wang, T. Hiroshima, A. Tomita, and M. Hayashi,
{\em Physics Reports} 448, 1(2007)
\bibitem{H03}
W.-Y.~Hwang, Phys. Rev. Lett. {\bf 91}, 057901 (2003).

\bibitem{wang05}
X.-B.~Wang, Phys. Rev. Lett. {\bf 94}, 230503 (2005); X.-B.~Wang,
Phys. Rev. A {\bf 72}, 012322 (2005).


\bibitem{LMC05}
H.-K.~Lo, X.~Ma, and K.~Chen, Phys. Rev. Lett. {\bf 94}, 230504
(2005); X.~Ma, B. Qi, Y. Zhao, and H.-K. Lo, Phys. Rev. A {\bf
72}, 012326 (2005).

\bibitem{HQph}
J.W.~Harrington {\em et al.}, quant-ph/0503002.
\bibitem{pdc1}W. Mauerer and C. Silberhorn, Phys. Rev. A {\bf 75} 050305 (2007);
Y. Adachi, T. Yamamoto, M. Koashi, and N. Imoto, Phys. Rev.Lett. {\bf 99}, 180503 (2008).
\bibitem{pdc2}
T. Hirikiri and T. Kobayashi, Phys. Rev. A (2006), 73, 032331; Q.
Wang, X.-B. Wang, G.-C. Guo, Phys. Rev. A (2007), 75, 012312.
\bibitem{haya}M. Hayashi, N. J. Phys., {\bf 9} 284.
\bibitem{zei} R. Ursin {\em et al.}, quant-ph/0607182.

\bibitem{scran}
V.~Scarani, A.~Acin, G.~Ribordy, N.~Gisin, Phys. Rev. Lett. 92,
057901 (2004); C.~Branciard, N.~Gisin, B.~Kraus, V.~Scarani, Phys.
Rev. A 72, 032301 (2005).

\bibitem{kko} M. Koashi, Phys. Rev. Lett., 93, 120501(2004);
K.~Tamaki, N.~L\"ukenhaus, M.~Loashi, J.~Batuwantudawe,
quant-ph/0608082



\bibitem{Lo06}
Y.~Zhao, B. Qi, X. Ma, H.-K. Lo and L. Qian, Phys. Rev. Lett. {\bf
96}, 070502 (2006); Y.~Zhao, B. Qi, X. Ma, H.-K. Lo and L. Qian,
quant-ph/0601168.
\bibitem{peng}  Cheng-Zhi Peng {\em et al.}
 Phys. Rev. Lett. 98, 010505 (2007); D. Rosenberg {em et al.},  {\em Phys. Rev. Lett.} 98, 010503
(2007),  T. Schmitt-Manderbach {\em et al.}, {\em Phys. Rev. Lett.}
98, 010504 (2007).
\bibitem{yuan} Z.-L. Yuan, A. W. Sharpe, and A. J. Shields,
{\em Appl. Phys. Lett.} 90, 011118 (2007).
\bibitem{wang07}X.-B. Wang, {\em Phys. Rev.} A75, 012301(2007)
\bibitem{wangapl}X.-B. Wang, C.-Z. Peng and J.-W. Pan, {\em Appl. Phys.
Let.} 90, 031110(2007)
\bibitem{note} Xiang-Bin Wang, C.-Z. Peng, J. Zhang, and Jian-Wei Pan,
quant-ph/0612121.
\bibitem{gisind}N. Gisin, S. Fasel, B. Kraus, H. Zbinden, and G.
Ribordy, Phys. Rev. A 73, 022320(2006).




\end{thebibliography}
\end{document}